\documentclass[conference,letterpaper]{IEEEtran}
\IEEEoverridecommandlockouts 
\usepackage[utf8]{inputenc} 
\usepackage[T1]{fontenc}
\usepackage{url}
\usepackage{ifthen}
\usepackage{cite}
\usepackage[cmex10]{amsmath} 
\usepackage{notation_style}
\usepackage{xcolor}         
\usepackage{subcaption}
\usepackage{caption}
\usepackage{subequation}

\begin{document}
\title{Leveraging Noisy Observations in Zero-Sum Games} 

%

 \author{%
	   \IEEEauthorblockN{Emmanouil M.~Athanasakos\IEEEauthorrefmark{1} and
		                     Samir M.~Perlaza\IEEEauthorrefmark{1}\IEEEauthorrefmark{2}\IEEEauthorrefmark{3}	                  
		                   \thanks{\noindent This work is supported by the Inria Exploratory Action–Information and Decision Making(AExIDEM).}}
	   \IEEEauthorblockA{\IEEEauthorrefmark{1}%
		                   INRIA, Centre Inria d’Université Côte d’Azur, Sophia Antipolis, France.}
	   \IEEEauthorblockA{\IEEEauthorrefmark{2}%
		                     ECE Dept. Princeton University, Princeton, 08544 NJ, USA.}
	   \IEEEauthorblockA{\IEEEauthorrefmark{3}%
		                     GAATI, Université de la Polynésie Française, Faaa, French Polynesia.}
		                     
	 }

\maketitle

\begin{abstract}
 This paper studies an instance of zero-sum games in which one player~(the leader) commits to its opponent~(the follower) to choose its actions by sampling a given probability measure~(strategy). The actions of the leader are observed by the follower as the output of an arbitrary channel. In response to that, the follower chooses its action based on its current information, that is, the leader's commitment and the corresponding noisy observation of its action. Within this context, the equilibrium of the game with noisy action observability is shown to always exist and the necessary conditions for its uniqueness are identified. Interestingly, the noisy observations have important impact on the cardinality of the follower's set of best responses. Under particular conditions, such a set of best responses is proved to be a singleton almost surely. The proposed model captures any channel noise with a density with respect to the Lebesgue measure. As an example, the case in which the channel is described by a Gaussian probability measure is investigated. 
\end{abstract}

\section{introduction}

Zero-sum games~(ZSGs) have proven to be a valuable tool across an array of disciplines, ranging from machine learning~(ML) to economics. ZSGs can be utilized to  proficiently model interactions between mutually adversarial decision makers~(players) and have taken on a pivotal role in the advancement of techniques that enhance the robustness of machine learning algorithms against adversarial attacks. Game theoretic adversarial learning/training techniques are introduced as a competitive game between the learner and the adversary in~\cite{Dalvi_04} and~\cite{Lowd_05}. Additionally, the fundamental concept of generative adversarial networks~(GANs) draws inspiration from a two-player zero-sum game played between the discriminator and the generator~\cite{goodfellow_2014_generative,goodfellow2014explaining,hsieh_2019_finding,oliehoek_2018_beyond}.\par 
The majority of the research at the intersection of game theory and adversarial ML utilizes the properties of ZSGs to forecast game results based on mixed strategies rather than individual actions. To put it simply, the core premise of the Stackelberg equilibrium~(SE)~\cite{Stackelberg-1952} in mixed strategies assumes that the leader's chosen strategy is completely observable by the follower, while actions remain unobservable. Conversely, in the context of the SE in pure strategies, the assumption is that the actions are perfectly observable, rendering the concept of committing to a strategy irrelevant~\cite{huang_2022_robust,bruckner_2011_stackelberg,zuo_2021_adversarial}. Regardless, these assumptions are rarely fulfilled in the majority of ML applications and can be challenging to substantiate their realism. Indeed, data is acquired through a data acquisition system, which is prone to quantization errors, additive noise, and distortions arising from data transmission and storage. These supplementary imperfections are not necessarily attributable solely to adversarial intent but rather stem from the inherent characteristics of transmission channels. This work makes a step towards this direction and studies new equilibrium concepts.
This work extends the game theoretic framework that was introduced in~\cite{sun_isit} by considering arbitrary channels. 
%
Driven by the growing relevance of game theory in the progress of ML~\cite{gao_2022_achieving,jin_2020_local,bai_2021_sample}, particularly in the context of GANs and adversarial learning, this work presents the subsequent contributions:

\begin{itemize}
	\item Under perfect observation of the commitment~(strategy of the leader), a new game formulation is introduced in which the follower observes a noisy version of actions played by the leader. In order to capture the impact of imperfect data acquisition, leader's actions are the inputs of an arbitrary channel, while the observations of the follower are the outputs of the channel.
	\item The formulation is not restricted to specific types of noise distributions; rather, it is general enough to facilitate any noise distribution that can be described by a probability density function~(pdf).
	\item The set of best responses of the follower is characterized and found that is a singleton almost surely. Moreover, the cases in which the noise is beneficial or irrelevant for the follower are investigated and the optimal commitments of the leader are characterized.
	\item The equilibrium is shown to always exist and the key conditions that affect its uniqueness are identified.
	\item An example for the case in which the channel is an additive white Gaussian noise channel~(AWGN) is investigated and the equilibrium solution is characterized.
\end{itemize}
\section{Preliminaries and Notation}\label{SecPrel}

\subsection{Game Formulation}\label{SecGameFormula}
Consider a two-player ZSG in normal form  in which \Pone and \Ptwo choose their actions from the sets $\set{A}_1$ and $\set{A}_2$, respectively. 
These sets are countable, and for all $k \inCountTwo$, $\set{A}_k \triangleq \{a_{k,1}, a_{k,2}, \ldots, a_{k,m_k}\}$ exhibits cardinality $m_k > 1$.
Let the function
\begin{IEEEeqnarray}{rcl}
	\label{EqMatrixU}
	u : \set{A}_1 \times  \set{A}_2 \to \reals,
\end{IEEEeqnarray}
be such that  if \Pone  plays $a_{1,i}$ and \Ptwo plays $a_{2,j}$, the payoff achieved by both players  is $u(a_{1,i},a_{2,j})$. Note that the function $u$ in~\eqref{EqMatrixU} is a simple function as its image is formed by the elements in the set $\{u(a_{1,i},a_{2,j}) : (i,j) \in  \setCountK{m_1} \times \setCountK{m_2}\}$. For the ease of notation, for all $(i,j) \in \setCountK{m_1} \times \setCountK{m_2}$ consider that $u_{i,j} \triangleq u(a_{1,i},a_{2,j})$. 
\Pone and \Ptwo choose their actions to maximize and minimize their payoffs, respectively.

When the game is played with commitments and one of the players observes the action played by its counterpart subject to noise, the notion of memoryless channels in the sense of~\cite{Shannon-1948a,Shannon-1948b} is utilized.  
%
Let $\simplex{\tilde{\set{A}_2}|\set{A}_2}$ be the set of all probability measures on $\tilde{\set{A}_2}$ conditioned on the elements of $\set{A}_2$. The channel is represented by $P_{\tilde{A}_2 | A_2} \in \simplex{\tilde{\set{A}_2}|\set{A}_2}$, for some given set $\tilde{\set{A}_2}$. More specifically, given a channel input $b \in \set{A}_2$, the channel output is $\tilde{b} \in \tilde{\set{A}_2}$ with probability $P_{\tilde{A}_2 | A_2 = b}(\tilde{b})$. A strategy for \Pone, denoted by $P_{A_1| \tilde{A}_2} \in \simplex{\set{A}_1|\tilde{\set{A}_2}}$, is a probability measure on $\set{A}_1$ conditioned on the elements of~$\tilde{\set{A}}_2$. 
%
%
Hence, when the channel output is $\tilde{b}$, \Pone chooses its actions by sampling the probability measure $P_{A_1| \tilde{A}_2 = \tilde{b}} \in \simplex{\set{A}_1}$, where $\simplex{\set{A_1}}$ denotes the set of all probability measures on the set $\set{A}_1$. 
Alternatively, a strategy for \Ptwo  is a probability measure denoted by $P_{A_2} \in \simplex{\set{A}_2}$. That is, \Ptwo chooses its actions by sampling $P_{A_2}$.

The extended ZSG that includes the noisy channel $P_{\tilde{A}_2 | A_2}$ is represented by the tuple:
\begin{IEEEeqnarray}{rcl}
	\label{EqTheNewGame}
	\gameNF{u,  P_{\tilde{A}_2 | A_2}} & \triangleq & \left( \set{A}_1 , \set{A}_2 , u, P_{\tilde{A}_2 | A_2} \right).
\end{IEEEeqnarray}
The game $\gameNF{u,  P_{\tilde{A}_2 | A_2}}$ is played in three stages.  
In the first stage, \Ptwo announces its strategy $P_{A_2} \in \simplex{\set{A}_2}$ to \Pone and commits to choose its actions by using such a strategy.
In the second stage, \Ptwo plays action $b \in \set{A}_2$ with probability $P_{A_2}(b)$ and \Pone observes the channel output $\tilde{b} \in \tilde{\set{A}}_2$ with probability $P_{\tilde{A}_2| A_2 = b}(\tilde{b})$.
In the final stage, \Pone plays action $a \in \set{A}_1$, with probability $P_{A_1| \tilde{A}_2 = \tilde{b}}\left( a \right)$ and both players obtain equal payoffs $u(a,b)$.

Therefore, the expected payoff obtained by the players is determined by the function $v: 
\simplex{\set{A}_1|\tilde{\set{A}_2}} \times \simplex{\set{A}_2} \to 
\reals$, such that given the strategy  $P_{A_1| \tilde{A}_2}$ of \Pone and the strategy $P_{A_2}$ of \Ptwo, the 
expected payoff is  
\begin{IEEEeqnarray}{rcl}
	\nonumber
	v\left(P_{A_1| \tilde{A}_2}, P_{A_2} \right) &=& \int_{\set{A}_2} \bigg( \int_{\tilde{\set{A}}_2} \left( \int_{\set{A}_1} u(a,b) \mathrm{d}P_{A_1| \tilde{A}_2=\tilde{b}}(a) \right)	\IEEEeqnarraynumspace \middlesqueezeequ \\
	\label{EqTheCostFunction} 
	& & \mathrm{d}P_{\tilde{A}_2| A_2=b}(\tilde{b}) \bigg) \mathrm{d}P_{A_2}(b).
\end{IEEEeqnarray}
While the sets $\set{A}_1$ and $\set{A}_2$ are assumed to be countable, the set $\tilde{\set{A}_2}$ is assumed to be a subset of $\mathbb{R}$. For all $j\in \{1,\ldots,m_2\}$, the channel $P_{\tilde{A}_2| A_2}$ satisfies
\begin{IEEEeqnarray}{rcl}
	\label{Eq_assumption_1}
	P_{\tilde{A}_2| A_2=a_{2,j}} \ll \gg \lambda,
\end{IEEEeqnarray}
where $\ll \gg $ denotes mutual absolute continuity; and $\lambda$ denotes the Lebesgue measure on  $(\mathbb{R}, \mathscr{B}(\mathbb{R}))$.

\noindent Under these assumptions, the expected payoff function $v$ in~\eqref{EqTheCostFunction} satisfies
\begin{IEEEeqnarray}{rcl}
	\nonumber
	v&&\left(P_{A_1| \tilde{A}_2}, P_{A_2} \right)  =   \int_{\tilde{\set{A}}_2} \Bigg(\sum_{i = 1}^{m_1} P_{A_1| \tilde{A}_2=\tilde{b}}(a_{1,i}) \\
\label{EqTheCostFunctionCountable_3} 
& &\bigg(\sum_{j = 1}^{m_2} u_{i,j}  P_{A_2}(a_{2,j}) 
\,\RND{P_{\tilde{A}_2| A_2=a_{2,j}}}{P_{\tilde{A}_2| A_2=a_{2,1}}}\,(\tilde{b})
\bigg) \Bigg) \mathrm{d}P_{\tilde{A}_2| A_2=a_{2,1}}(\tilde{b}), \IEEEeqnarraynumspace \Tsupersqueezeequ
\end{IEEEeqnarray}
where for all $j \in \{1,\ldots,m_2\}$, 
the function $\RND{P_{\tilde{A}_2| A_2=a_{2,j}}}{P_{\tilde{A}_2| A_2=a_{2,1}}} : \tilde{\set{A}_2}  \to (0,+\infty) $ is the Radon-Nikodym derivative of $P_{\tilde{A}_2| A_2=a_{2,j}}$ with respect to $P_{\tilde{A}_2| A_2=a_{2,1}}$.

\subsection{Equilibrium}\label{subsec_equilib}
The necessary mathematical objects for characterizing the solution concept of the extended game $\gameNF{u,  P_{\tilde{A}_2 | A_2}}$ in~\eqref{EqTheNewGame} are defined in this section.
\begin{definition}\label{Def_Best_response_P1}
	\emph{(Best Responses of \Pone) Given a commitment $P \in \simplex{\set{A}_2}$ and a channel output $\tilde{b}\in \tilde{\set{A}_2}$, the set of best responses of Player 1, is determined by the correspondence $\BR_1: \simplex{\set{A}_{2}} \times \tilde{\set{A}_2}\to \mathscr{P}\left(\simplex{\set{A}_1}\right)$, where $\mathscr{P}\left(\simplex{\set{A}_1}\right)$ denotes the power set of $\set{A}_1$, such that
		\begin{IEEEeqnarray}{rcl}
			\nonumber
			\BR_1( P, \tilde{b} ) & = & \arg\max_{ Q \in \simplex{\set{A}_1}} \sum_{i = 1}^{m_1} Q (a_{1,i}) \bigg(\sum_{j = 1}^{m_2} u_{i,j}  P(a_{2,j}) \IEEEeqnarraynumspace \middlesqueezeequ \\
			\label{EqBR1_1}		
			& &\RND{P_{\tilde{A}_2| A_2=a_{2,j}}}{P_{\tilde{A}_2| A_2=a_{2,1}}}(\tilde{b})
			\bigg).  \IEEEeqnarraynumspace \middlesqueezeequ
		\end{IEEEeqnarray}  
		A strategy $P_{A_1| \tilde{A}_2} \in \simplex{\set{A}_1|\tilde{\set{A}_2}}$ of \Pone is a best response to the commitment $P \in \simplex{\set{A}_2}$, if for all $\tilde{b} \in \tilde{\set{A}_2}$ the probability measure $P_{A_1| \tilde{A}_2 = \tilde{b}} \in \BR_1( P, \tilde{b} )$. }
\end{definition}
\noindent Let the function $\hat{v}: \simplex{\set{A}_2} \to \mathbb{R}$ be such that for a given commitment $P \in \simplex{\set{A}_2}$,
\begin{IEEEeqnarray}{rcl}
	\label{Eq_Best_PLayer2}
	\hat{v}(P) = \max_{Q_{A_1| \tilde{A}_2} \in \simplex{\set{A}_1|\tilde{\set{A}_2}}} v(Q_{A_1| \tilde{A}_2}, P), \\
	\nonumber
	\text{s.t.}\, \forall \tilde{b} \in \tilde{\set{A}_2} \quad Q_{A_1| \tilde{A}_2=\tilde{b}} \in \BR_1( P, \tilde{b} ), 
\end{IEEEeqnarray}
where the function $v$ is defined in \eqref{EqTheCostFunctionCountable_3} and the correspondence $\BR_1$ in~\eqref{EqBR1_1}.

\noindent Accordingly, the set of optimal commitments is the set of minimizers of the function $\hat{v}$ given in~\eqref{Eq_Best_PLayer2}. This observation leads to the following notion of equilibrium.
\begin{definition}\label{Def_Equlibrium}
	\emph{(Equilibrium) 
		The strategies $P^{\dagger}_{A_1| \tilde{A}_2} \in \simplex{\set{A}_1|\tilde{\set{A}_2}}$ and $P^{\dagger}_{A_2} \in \simplex{\set{A}_2}$ for \Pone and \Ptwo, respectively,
		are said to form an equilibrium of the game $\gameNF{u,  P_{\tilde{A}_2 | A_2}}$ in \eqref{EqTheNewGame} if
		\begin{IEEEeqnarray}{rcl}
			\label{Eq_equilibrium_1}
			P^{\dagger}_{A_2} \in \arg \min_{P \in \simplex{\set{A}_2} } \hat{v}(P)	
		\end{IEEEeqnarray} 
		and for all $\tilde{b} \in \tilde{\set{A}_2}	$
		\begin{IEEEeqnarray}{rcl}	 
			\label{Eq_equilibrium_2}
			P^{\dagger}_{A_1|\tilde{A}_2 = \tilde{b}} \in  \BR_1(P^{\dagger}_{A_2}, \tilde{b} ), 		
		\end{IEEEeqnarray}	
		where the function $\hat{v}$ is defined in \eqref{Eq_Best_PLayer2} and the correspondence $\BR_1$ is defined in~\eqref{EqBR1_1}.}
\end{definition}
\subsection{Existing Special Cases of Game $\gameNF{u,  P_{\tilde{A}_2 | A_2}}$}
%

Consider that the channel $P_{\tilde{A}_2 | A_2}$ is such that for all $\tilde{b} \in \tilde{\set{A}_2}$, $P_{\tilde{A}_2 | A_2=a_{2,j}}(\tilde{b}) = P_{\tilde{A}_2 | A_2=a_{2,r}}(\tilde{b})$, with $(j,r)=\{1,\ldots,m_2\}^2$ and $j\neq r$. This is the case in which the channel input and output are independent. In such case, the game $\gameNF{u, P_{\tilde{A}_2 | A_2}}$ simplifies to the game denoted by
\begin{IEEEeqnarray}{rcl}
	\label{EqTheGame}
	\gameNF{u} & \triangleq & \left( \set{A}_1 , \set{A}_2 , u \right).
\end{IEEEeqnarray} 
When players simultaneously choose their actions in the absence of commitment the game in~\eqref{EqTheGame} has been studied in~\cite{von_1947_theory}. The solution concept is the Nash Equilibrium~(NE), as introduced in~\cite{nash_1950_equilibrium}.\par 

When there exist a deterministic one-to-one mapping between the inputs and the outputs of the channel, the game $\gameNF{u, P_{\tilde{A}_2 | A_2}}$ is reduced to the game studied in~\cite{von1934marktform} and~\cite{simaan_1973_additional},~\cite{simaan_1973_stackelberg}. This is the case in which for all $a_{2,j} \in \set{A}_2$, with $j=\{1,\ldots,m_2\}$, it holds that $P_{\tilde{A}_2 | A_2=a_{2,j}}(\tilde{b}) = 1$. That is, regardless the commitment, \Pone can always select an optimal action in response to \Ptwo's action.\par 
When the commitment is the only information that \Pone has regarding the action played by \Ptwo, \ie, the channel input and output are independent, the game $\gameNF{u,  P_{\tilde{A}_2 | A_2}}$ is equivalent to the one investigated in~\cite{conitzer_2016_stackelberg} and~\cite{leonardos_2018_commitment}, and the solution concept is the SE in mixed strategies.\par 
Other special cases of the game $\gameNF{u,  P_{\tilde{A}_2 | A_2}}$ are studied in~\cite{bagwell_1995_commitment} and~\cite{sun_isit}. In the former, the strategies of the players are restricted to pure strategies and the sets $\set{A}_2$ and $\tilde{\set{A}_2}$ are finite and assumed to be identical. In the latter, the analysis is generalized to include the consideration of mixed strategies.


%
\section{main results}\label{main_res}
%

%
\noindent For the rest of this work, it is assumed that
%
			for all $(i,l) \in \{1,\ldots,m_1\}^2$,
			\begin{IEEEeqnarray}{rcl}
\label{Eq_assumption_2}	
				\lambda\left(\hspace{-0.9ex}\biggl\{\tilde{b} \in \tilde{\set{A}_2} \,:\, \sum_{j = 1}^{m_2} \left(u_{i,j}-u_{l,j}\right)  P(a_{2,j}) f_{\tilde{A}_2| A_2=a_{2,j}}(\tilde{b}) = 0\biggr\}\hspace{-0.9ex}\right)=0, \IEEEeqnarraynumspace\Tsupersqueezeequ 										
			\end{IEEEeqnarray}
			where $f_{\tilde{A}_2| A_2=a_{2,j}}$ is the corresponding pdf of the measure $P_{\tilde{A}_2| A_2=a_{2,j}}$. 	

The assumption in~\eqref{Eq_assumption_2} is always satisfied when for all $(i,l) \in \{1,\ldots,m_1\}^2$, it holds that $ \prod_{j=1}^{m_2}\left(u_{i,j}-u_{l,j}\right)  P(a_{2,j})>0$. In~\cite{sun_isit}, it has been shown that imposing such condition restricts the game $\gameNF{u}$ to exhibit a unique NE in mixed strategies.

%
The following lemma describes the set of best responses of \Pone.

\begin{lemma}\label{Lemma_Best_response_P1}
	Given a commitment $P \in \simplex{\set{A}_2}$ and a channel output $\tilde{b} \in \tilde{\set{A}_2}$, the set of best responses of \Pone is
	\begin{IEEEeqnarray}{rcl}
		\label{EqBR1_2}
		\BR_1( P, \tilde{b} )  \triangleq  \{ Q \in \simplex{\set{A}_1} : Q(a)>0 \text{ if } a \in \set{S}(P,\tilde{b})   \},  \IEEEeqnarraynumspace \middlesqueezeequ
	\end{IEEEeqnarray}
	where
	\begin{IEEEeqnarray}{rcl}
		\label{EqSetBR1}
		\set{S}(P,\tilde{b}) \,\triangleq\, \arg \max_{a \in \set{A}_1} \bigg\{\hspace{-0.3ex}\sum_{j = 1}^{m_2} u(a, a_{2,j})  P(a_{2,j})\, \RND{P_{\tilde{A}_2| A_2=a_{2,j}}}{P_{\tilde{A}_2| A_2=a_{2,1}}}\,(\tilde{b}) \hspace{-0.3ex}\bigg\}.\IEEEeqnarraynumspace \Tsupersqueezeequ
	\end{IEEEeqnarray}
\end{lemma}
\begin{IEEEproof}
	The proof of is presented in Appendix A of~\cite{ath_inria_report}.
\end{IEEEproof}
The preceding lemma indicates the fact that given a commitment $P \in \simplex{\set{A}_2}$ and a channel output $\tilde{b} \in \tilde{\set{A}_2}$, \Pone's optimal strategy is to concentrate the probability measure $P_{A_1| \tilde{A}_2=\tilde{b}}$ over the set $\set{S}(P,\tilde{b}) \subseteq \set{A}_1$, with $\set{S}(P,\tilde{b})$ in~\eqref{EqSetBR1}. If $|\set{S}(P,\tilde{b})| = 1$, there is a unique best response $P^{\star}_{A_1| \tilde{A}_2=\tilde{b}} \in \BR_1( P, \tilde{b} )$ that assigns probability one to the singleton $\set{S}(P,\tilde{b})$. The set of best responses of \Pone might also contain infinitely many strategies. In the specific case in which $\BR_1( P, \tilde{b} ) = \simplex{\set{A}_1}$, then as observed in~\cite{sun_isit}, \Pone opts for indifferently choosing its actions. Interestingly, under assumptions~\eqref{Eq_assumption_1} and~\eqref{Eq_assumption_2}, for each channel output, \Pone deterministically chooses a specific action almost surely, which will be discussed in the sequel.

%
For all $i \in \{1,\ldots,m_1\}$ and for all $P \in \simplex{\set{A}_2}$, let the set $\set{H}_i(P)$ be
\begin{IEEEeqnarray}{rcl}
	\nonumber
	\set{H}_i(P) \triangleq \biggl\{\tilde{b} \in \tilde{\set{A}_2} &\,:\,& i \in \arg \max \bigg\{\sum_{j = 1}^{m_2} u_{i,j}  P(a_{2,j})  \IEEEeqnarraynumspace \Tsupersqueezeequ \\ 
	\label{partition_1}
	&&\RND{P_{\tilde{A}_2| A_2=a_{2,j}}}{P_{\tilde{A}_2| A_2=a_{2,1}}}\,(\tilde{b}) \bigg\}\biggr\} \mathbin{\Big\backslash} \set{H}_{i-1}(P),  \IEEEeqnarraynumspace \Tsupersqueezeequ
\end{IEEEeqnarray}
with $\set{H}_{0}(P) \triangleq \emptyset $. Let also for all $(i,l) \in \{1,\ldots,m_1\}^2$, with $i \neq l$, the set $\set{H}_{i,l}$ be
\begin{IEEEeqnarray}{rcl}
\nonumber
	\set{H}_{i,l}(P) \triangleq \biggl\{\tilde{b} \in \tilde{\set{A}_2} &:& \sum_{j = 1}^{m_2} u_{i,j}  P(a_{2,j}) \RND{P_{\tilde{A}_2| A_2=a_{2,j}}}{P_{\tilde{A}_2| A_2=a_{2,k}}}(\tilde{b}) \\
		\label{partition_2}
	&=& \sum_{j = 1}^{m_2} u_{l,j}  P(a_{2,j}) \RND{P_{\tilde{A}_2| A_2=a_{2,j}}}{P_{\tilde{A}_2| A_2=a_{2,1}}}\,(\tilde{b}) \biggr\}.\IEEEeqnarraynumspace \Tsupersqueezeequ
\end{IEEEeqnarray}
Note that $\set{H}_{1}(P), \set{H}_{2}(P), \ldots, \set{H}_{m_1}(P)$ form a partition of $\tilde{\set{A}_2}$. The following theorem presents a property exhibited by the set $\set{H}_{i,l}(P)$ in~\eqref{partition_2}.
%
%
%
\begin{theorem}\label{theorem_zero_measure}
	Given a commitment $P \in \simplex{\set{A}_2}$, under assumptions~\eqref{Eq_assumption_1} and~\eqref{Eq_assumption_2}, if for all $(i,l) \in \{1,\ldots,m_1\}^2$, with $i \neq l$, and for all $a_{2,j}\in \set{A}_2$
	\begin{IEEEeqnarray}{rcl}
		\label{Eq_condition_nonzero_util}
		\left(u_{i,j}-u_{l,j}\right) P(a_{2,j}) \neq 0,
	\end{IEEEeqnarray}
then it holds that	
	\begin{IEEEeqnarray}{rcl}
		\label{eq_zero_measure}
		P_{\tilde{A}_2}\left(\set{H}_{i,l}(P)\right)= 0,
	\end{IEEEeqnarray}
	where $\set{H}_{i,l}(P)$ is defined in~\eqref{partition_2}; and the probability measure $P_{\tilde{A}_2}$ is such that for all measurable subsets $\set{A}$ of $\tilde{\set{A}_2}$, it holds that
	\begin{IEEEeqnarray}{rcl}
		\label{eq_zero_measure_measure_P}
		P_{\tilde{A}_2}(\set{A}) = \int P_{\tilde{A}_2| A_2=a}(\set{A})\mathrm{d}P(a).
	\end{IEEEeqnarray}
\end{theorem}
\begin{IEEEproof}
	The proof of is presented in Appendix B of~\cite{ath_inria_report}.
\end{IEEEproof}
\noindent A first observation from Theorem~\ref{theorem_zero_measure}, is that if \Ptwo's commitment is in strict mixed strategies, then the elements of the set $\set{H}_{i,l}(P)$ are observed with probability zero with respect to $P_{\tilde{A}_2}$ in~\eqref{eq_zero_measure_measure_P}. Note that an element $\tilde{b}$ of the set $\set{H}_{i,l}(P)$ is the output of the channel for which actions $a_{1,i}$ and $a_{1,l}$ might be both best responses for \Pone to the commitment $P$ and the channel output $\tilde{b}$. Hence, under the assumptions of Theorem~\ref{theorem_zero_measure}, the set $\set{S}(P,\tilde{b})$ is always a singleton and thus $|\BR_1(P, \tilde{b})| = 1$. 
On the other hand, if~\eqref{Eq_condition_nonzero_util} does not hold, then the set of best responses of \Pone includes infinitely many strategies. This indicates that \Pone indifferently chooses its actions from the set $\set{S}(P,\tilde{b})$.
This is formalized by the following lemma.

\begin{lemma}\label{Lemma_Best_response_P1_strategy}
	Under assumptions~\eqref{Eq_assumption_1} and~\eqref{Eq_assumption_2}, given a commitment $P \in \simplex{\set{A}_2}$ that satisfies~\eqref{Eq_condition_nonzero_util}, for all $\tilde{b} \in \tilde{\set{A}_2}$ and for all $P_{A_1| \tilde{A}_2} \in \BR_1( P, \tilde{b})$, with $\BR_1( P, \tilde{b})$ in~\eqref{EqBR1_1}, it holds that for all $i \in \{1,\ldots,m_1\}$
	\begin{IEEEeqnarray}{rcl}
		\label{Eq_BR_general}
		P_{A_1| \tilde{A}_2 = \tilde{b}}(a_{1,i}) = 
		\begin{cases} 
			1, \text{ if } \tilde{b} \in \set{H}_i(P) \\ 
			0, \text{ otherwise}, 
		\end{cases}
	\end{IEEEeqnarray}
	where $\set{H}_i(P)$ is defined in~\eqref{partition_1} and $a_{1,i} \in \set{A}_1$. Moreover, the set of best responses satisfies
	\begin{IEEEeqnarray}{rcl}
		\label{Eq_BR_singleton_gen}
		P_{\tilde{A}_2} \left( \biggl\{\tilde{b} \in \tilde{\set{A}_2} : \left|\BR_1\left( P, \tilde{b} \right)\right| = 1\biggr\} \right) = 1.
	\end{IEEEeqnarray}
	%
\end{lemma}
\begin{IEEEproof}
	The proof of is presented in Appendix C of~\cite{ath_inria_report}.
\end{IEEEproof}
%
%
%
%
%
%


\noindent The following theorem characterizes the equilibrium of the extended game $\gameNF{u,  P_{\tilde{A}_2 | A_2}}$ in~\eqref{EqTheNewGame}.
\begin{theorem}\label{Th_Gauss_Existence}			
	 Under assumptions~\eqref{Eq_assumption_1} and~\eqref{Eq_assumption_2}, and if~\eqref{Eq_condition_nonzero_util} holds for all $j \in \{1,\ldots,m_2\}$ and for all $(i,l) \in \{1,\ldots,m_1\}^2$, with $i \neq l$, then the game $\gameNF{u, P_{\tilde{A}_2 | A_2}}$ possesses a unique equilibrium almost surely with respect to $P_{\tilde{A}_2}$.
\end{theorem}
\begin{IEEEproof}
	The proof is presented in Appendix D of~\cite{ath_inria_report}.
\end{IEEEproof}
The game $\gameNF{u}$ in~\eqref{EqTheGame} holds significant importance due to the fact that its the expected payoff can be used as reference point when analyzing the game in~\eqref{EqTheNewGame}. Let the function $\omega: \simplex{\set{A}_1} \times \simplex{\set{A}_2} \to \mathbb{R}$ represent the expected payoff of game $\gameNF{u}$ such that
\begin{IEEEeqnarray}{rcl}
	\label{Eq_payoff_normal}
	\omega(P_{A_1}, P_{A_2}) =\sum_{i = 1}^{m_1} \sum_{j = 1}^{m_2} P_{A_1}(a_{1,i}) u_{i,j} P_{A_2}(a_{2,j}).
\end{IEEEeqnarray}

\noindent Let the function $\hat{u}: \simplex{\set{A}_2} \to \mathbb{R}$ be such that for all commitments $P \in \simplex{\set{A}_2}$,
\begin{IEEEeqnarray}{rcl}
	\label{Eq_fun_u_hat}
	\hat{u}(P) = \max_{Q \in \simplex{\set{A}_1}} \omega(Q, P),
\end{IEEEeqnarray}
where the function $\omega$ is defined in~\eqref{Eq_payoff_normal}.
The term $\hat{u}(P)$ in~\eqref{Eq_fun_u_hat} represents the expected payoff of the game $\gameNF{u}$ in~\eqref{EqTheGame}, when \Ptwo commits to play strategy $P$. Note that, the minimum value over all commitments $P$ of the function $\hat{u}$ is the payoff at the NE. The following lemma compares the payoffs at the equilibria of the games $\gameNF{u}$ in~\eqref{EqTheGame} and $\gameNF{u, P_{\tilde{A}_2 | A_2}}$ in~\eqref{EqTheNewGame}.
\begin{lemma}\label{lemma_inequality_comparison}
	 Let the probability measures $\left(P^{\dagger}_{A_1| \tilde{A}_2}, P^{\dagger}_{A_2}\right)$ form an equilibrium of the game $\gameNF{u,  P_{\tilde{A}_2 | A_2}}$ in~\eqref{EqTheNewGame} and let the pair of strategies $\left(P_{A_1}^{\star}, P_{A_2}^{\star}\right) \in \simplex{\set{A}_1} \times \simplex{\set{A}_2}$ be an NE of the game $\gameNF{u}$ in~\eqref{EqTheGame}. Then for all commitments $P \in \simplex{\set{A}_2}$, it holds that 
	\begin{IEEEeqnarray}{rcl}
		\label{Eq_inequality_compare}
		\hat{u}(P_{A_2}^{\star}) \leq v\left(P^{\dagger}_{A_1| \tilde{A}_2}, P^{\dagger}_{A_2}\right)\leq \hat{v}(P) \leq \min_{j} \max_{i} u_{i,j},
	\end{IEEEeqnarray}
	where the functions $\hat{u}$, $\hat{v}$ and $v$ are given in~\eqref{Eq_fun_u_hat},~\eqref{Eq_Best_PLayer2} and~\eqref{EqTheCostFunctionCountable_3}, respectively.	
\end{lemma}
\begin{IEEEproof}
	The proof is presented in Appendix E of~\cite{ath_inria_report}.	
\end{IEEEproof}

\noindent Lemma~\ref{lemma_inequality_comparison} unveils the fact that the payoff of the game $\gameNF{u, P_{\tilde{A}_2 | A_2}}$ at the equilibrium is lower bounded by the payoff of the game $\gameNF{u}$ at the NE and upper bounded by the payoff of such a game at the SE in pure strategies. The former is the case in which \Pone does not observe the actions of \Ptwo~(\eg very noisy channel, $I(P_{\tilde{A}_2 | A_2}; P_{A_2}) = 0$); and the latter corresponds to the case in which \Pone perfectly observes the actions of its opponent~(\eg channel with zero noise, $I(P_{\tilde{A}_2 | A_2}; P_{A_2}) = H(P_{A_2})$).
%

%
\section{Example: Gaussian channel}\label{awgn_sec}
Thus far, the probability measure that describes the channel $P_{\tilde{A}_2 | A_2}$ of the game $\gameNF{u,  P_{\tilde{A}_2 | A_2}}$ in~\eqref{EqTheNewGame} is not yet specified. This generality signifies a key feature of the theoretical model, which captures several data processing impairments. Moreover, while the specific game solution concept relies on the characteristics of the channel, the generality of the game formulation presented in Section~\ref{SecPrel} highlights its capability to effectively address any type of noise which has a density with respect the Lebesgue measure. In this section, the channel is assumed to be an AWGN channel and the action sets of \Pone and \Ptwo have the same cardinality and contain two actions. In this scenario, the channel is such that for all $a_{2,j} \in \set{A}_2$ and all measurable subsets $\set{A}$ of $\tilde{\set{A}_2}$
\begin{IEEEeqnarray}{rcl}
\nonumber
	P_{\tilde{A}_2 | A_2 = a_{2,j}} (\set{A}) =\, \frac{1}{\sqrt{2 \pi \sigma_j^2}} \int_{\set{A}} \exp\left(\frac{1}{2 \sigma_j^2} (\tilde{b}-\mu_j-a_{2,j})^2 \right) \mathrm{d}\lambda(\tilde{b}),
	 \Tsupersqueezeequ \\
	 	\label{Eq_Gaussian_measures_def1}
\end{IEEEeqnarray}
where $\lambda$ is the Lebesgue measure in $\mathbb{R}$; $\mu_j - a_{2,j}$ and $\sigma^2_j$ are the mean and the variance, respectively. 
From~\eqref{Eq_Gaussian_measures_def1}, it holds that the Radon-Nikodym derivative of $P_{\tilde{A}_2| A_2=a_{2,j}}$ with respect to $P_{\tilde{A}_2| A_2=a_{2,1}}$, and for all $\tilde{b} \in \tilde{\set{A}_2}$ is 
\begin{IEEEeqnarray}{rcl}
	\label{RND_gaussian}
	\RND{P_{\tilde{A}_2| A_2=a_{2,2}}}{P_{\tilde{A}_2| A_2=a_{2,1}}}\,(\tilde{b}) = \exp\left( - \left(\hspace{-0.3ex}\frac{(a_{2,2}^2 - a_{2,1}^2)}{2 \sigma^2}\ +\ \frac{\tilde{b}(a_{2,1}-a_{2,2})}{\sigma^2} \right) \hspace{-0.3ex}\right),\IEEEeqnarraynumspace \Tsupersqueezeequ 
\end{IEEEeqnarray}	
where it is assumed that $\mu_j=0$ and $\sigma_j = \sigma >0$, with $j \in \{1,2\}$. From~\eqref{partition_2}, for all $P_{A_2} \in \simplex{\set{A_2}}$ the set $\set{H}_{i,l}(P_{A_2})$, with $(i,l) \in \{1,2\}^2$, satisfies
\begin{IEEEeqnarray}{rcl}
	\nonumber
	\set{H}_{1,2}(P_{A_2}) &\triangleq& \left\lbrace\hspace{-0.3ex} \tilde{b} \in \tilde{\set{A}_2} :\, \RND{P_{\tilde{A}_2| A_2=a_{2,2}}}{P_{\tilde{A}_2| A_2=a_{2,1}}}(\tilde{b}) \,= \,\frac{\left(u_{1,1}-u_{2,1}\right) P_{A_2}(a_{2,1})}{\left(u_{2,2}-u_{1,2}\right)P_{A_2}(a_{2,2})}\hspace{-0.3ex}\right\rbrace		
	 \Tsupersqueezeequ \\
	 \label{Eq_hypothesis_0}\\
	 \nonumber
	 &=& \Biggl\{ \frac{\sigma^2}{a_{2,1}-a_{2,2}} \bigg(\log \left(\frac{\left(u_{1,1}-u_{2,1}\right) P_{A_2}(a_{2,1})}{\left(u_{2,2}-u_{1,2}\right) P_{A_2}(a_{2,2})}\right) \\
	\label{Eq_hypothesis_5}	 
	  &+& \mathrm{D}\left(P_{\tilde{A}_2| A_2=a_{2,2}} || P_{\tilde{A}_2| A_2=a_{2,1}}\right)\bigg) \Biggr\}, \Dsupersqueezeequ 
\end{IEEEeqnarray}	
where $\mathrm{D}$ denotes the Kullback-Leibler divergence. Note that the set $\set{H}_{1,2}(P_{A_2})$ is a singleton.
This said, it holds that $\lambda\left(\set{H}_{1,2}(P_{A_2})\right) = 0$ and thus, $P_{\tilde{A}_2}\left(\set{H}_{1,2}(P_{A_2})\right)=0$, which demonstrates the validity of Theorem~\ref{theorem_zero_measure}. 
%

Let the complementary cumulative distribution function~(ccdf) of the Gaussian pdf be
\begin{IEEEeqnarray}{rcl}
	\label{Q_gaussian}
	\mathrm{Q}\left(x\right) = \frac{1}{\sqrt{2\pi}} \int_{x}^{+\infty} \exp \left(\frac{-z^2}{2}\right)\mathrm{d}z.\IEEEeqnarraynumspace 	\supersqueezeequ	
\end{IEEEeqnarray}	
Using this notation, the following lemma provides an explicit expression for the function $\hat{v}$ in~\eqref{Eq_Best_PLayer2}
\begin{lemma}\label{Lemma_Gauss_v_hat}
	Given a commitment $P_{A_2} \in \simplex{\set{A}_2}$, if for all $a_{2,j} \in \set{A}_2$, the probability measure $P_{\tilde{A}_2 | A_2 = a_{2,j}}$ satisfies~\eqref{Eq_Gaussian_measures_def1}, then it holds
	\begin{IEEEeqnarray}{rcl}		
		\nonumber
		&&\hat{v}(P_{A_2}) = \max_{ Q_{A_1| \tilde{A}_2} \in \BR_1( P_{A_2}, \tilde{b})} v(Q_{A_1| \tilde{A}_2}, P_{A_2}) \IEEEeqnarraynumspace \supersqueezeequ\\
		\label{Eq_final_v_hat_Gauss}		
		&=& \sum_{j = 1}^{m_2}  P_{A_2}(a_{2,j})\bigg(u_{2,j}+ \left(u_{1,j}-u_{2,j}\right)\mathrm{Q}\left(\frac{b^{\star}_{P_{A_2}} - a_{2,j}}{\sigma}\right)\bigg), \IEEEeqnarraynumspace 	\Tsupersqueezeequ	
	\end{IEEEeqnarray}	
	where function $v$ is defined in~\eqref{EqTheCostFunctionCountable_3}; $b^{\star}_{P_{A_2}} \in \set{H}_{1,2}(P_{A_2})$; and function $\mathrm{Q}$ is in~\eqref{Q_gaussian}.
\end{lemma}
\begin{IEEEproof}
	The proof is presented in Appendix F of~\cite{ath_inria_report}.
\end{IEEEproof}
%
%

\begin{figure}[h]
	\centering
	\begin{subfigure}[b]{\linewidth}
		\includegraphics[width=\linewidth]{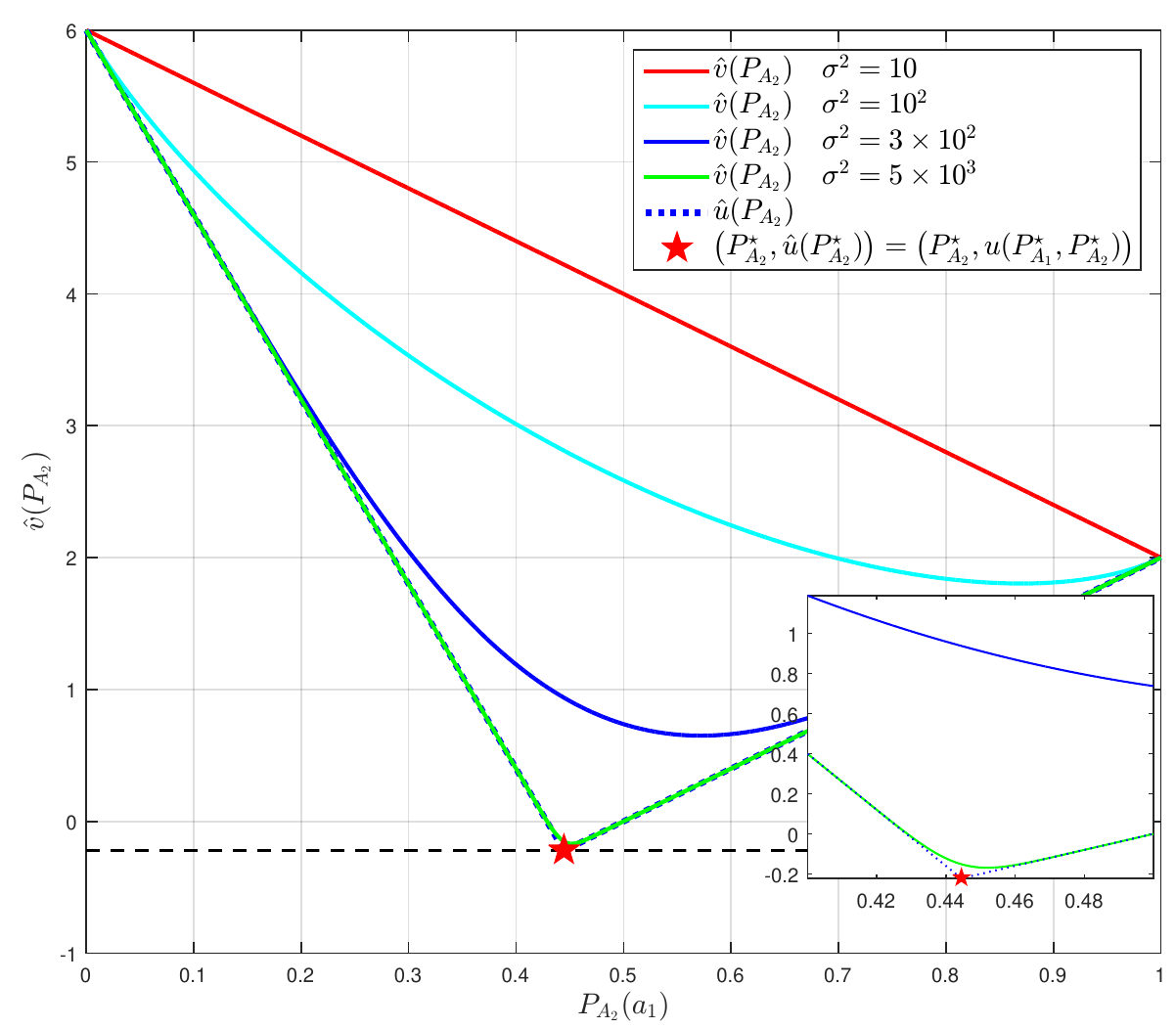}
		\caption{Payoff matrix $u =\left( 8, -6 ; -2, 2\right)$.}
		\label{fig_game_1} 
	\end{subfigure}
	\begin{subfigure}[b]{\linewidth}
		\includegraphics[width=\linewidth]{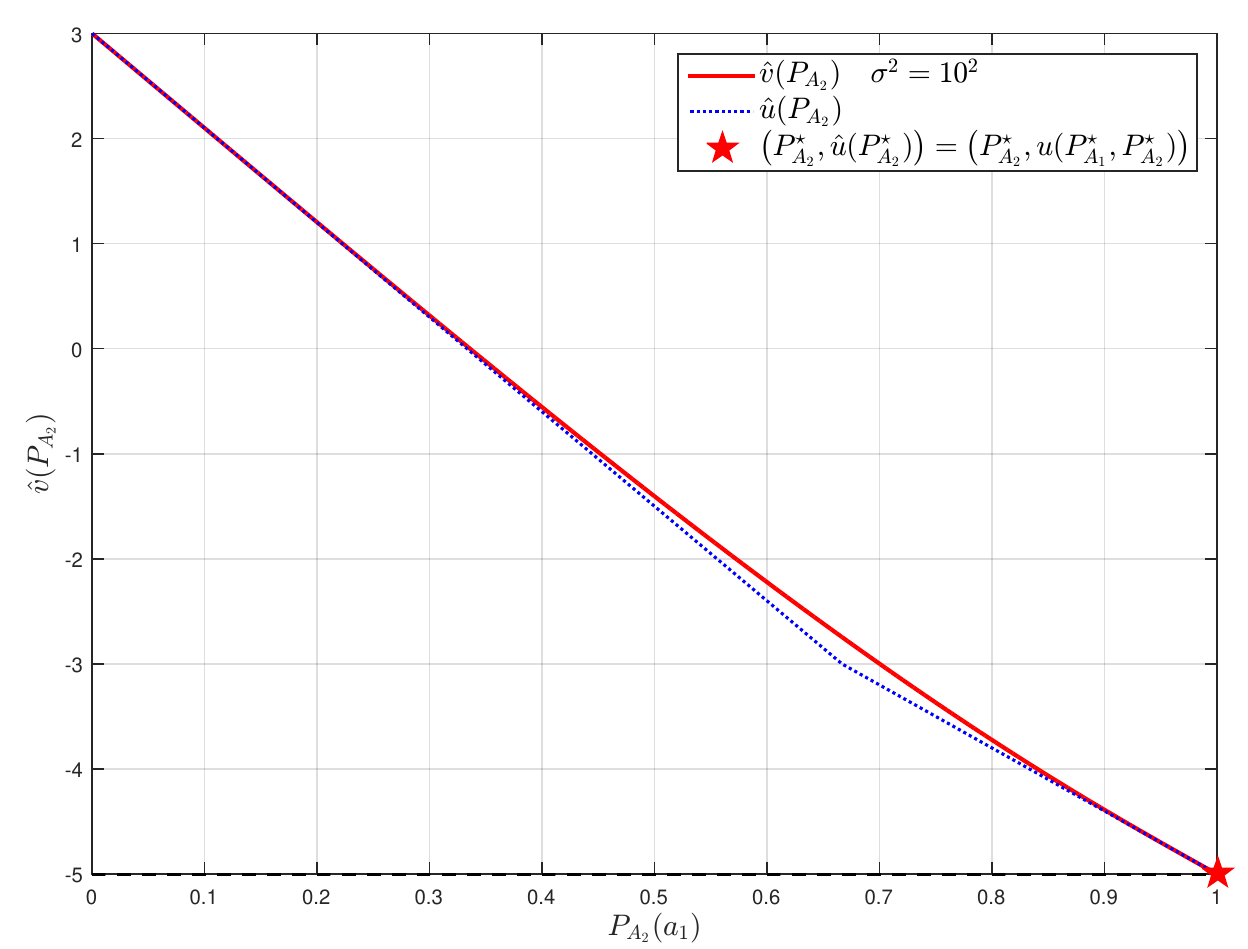}
		\caption{ Payoff matrix $u =\left( -5, 1 ; -6, 3\right)$.}
		\label{fig_game_2}
	\end{subfigure}
	\caption{ Plots of  $\hat{v}$ in~\eqref{Eq_Best_PLayer2} and $\hat{u}$ in~\eqref{Eq_fun_u_hat} as a function of the commitment $P_{A_2}$. The channel is the AWGN defined in~\eqref{Eq_Gaussian_measures_def1} with variance $\sigma^2_1=\sigma^2_2=\sigma^2$;  and $-a_{2,1} =a_{2,2}=10^2$.}
\end{figure}

In Fig.~\ref{fig_game_1}, the strategies $\left(P_{A_1}^{\star}, P_{A_2}^{\star}\right)$ with  $P_{A_1}^{\star}(a_{1,1}) = 1 - P_{A_1}^{\star}(a_{1,2})= 0.23$ and $P_{A_2}^{\star}(a_{2,1}) =1 - P_{A_21}^{\star}(a_{2,2})= 0.45$ form the unique NE of the game $\gameNF{u}$ whose payoff is $\hat{u}(P_{A_2}^{\star})$~(red star). Examining the behavior of the function $\hat{v}$ in~\eqref{Eq_Best_PLayer2} for a range of noise variance values $\sigma^2$, it is observed that; $(i)$ the equilibrium of the game $\gameNF{u,  P_{\tilde{A}_2 | A_2}}$ is always lower bounded by the equilibrium of $\hat{u}$~(dashed blue line in the zoomed subfigure) as Lemma~\ref{lemma_inequality_comparison} revealed; $(ii)$ the function $\hat{v}$ is monotonically decreasing with respect to $\sigma^2$; $(iii)$ letting $\sigma^2$ grow to infinity, results in the equivalence of the equilibrium of $\gameNF{u,  P_{\tilde{A}_2 | A_2}}$ and the NE of game $\gameNF{u}$~(green curve approaching the red star); and $(iv)$ letting $\sigma^2$ to be close to zero, the equilibrium payoff of the game $\gameNF{u, P_{\tilde{A}_2 | A_2}}$ in~\eqref{EqTheNewGame} becomes identical to the SE in pure strategies of the game $\gameNF{u}$ in~\eqref{EqTheGame} which is the min-max solution in pure strategies~(red line).
%

In Fig.~\ref{fig_game_2}, the game $\gameNF{u}$ exhibits a unique equilibrium in pure strategies and the NE payoff of the game $\gameNF{u}$ is $\hat{u}(P_{A_2}^{\star})$~(red star). Here, the equilibrium payoffs of both games are identical. In both examples it holds that $\hat{u}\left(P_{A_2}^{\star}\right) \leq v \left(P^{\dagger}_{A_1| \tilde{A}_2}, P^{\dagger}_{A_2}\right)$, which signifies that even subject to noise, the equilibrium payoff increases when \Pone observes a noisy version of the action played by \Ptwo.
%
%

%

\IEEEtriggeratref{14}
\bibliographystyle{ieeetr}
\bibliography{myreferences}

\end{document}